\documentclass[preprint,onecolumn,aps,pre,eqsecnum,longbibliography,showpacs,showkeys,a4paper]{revtex4-1}
\usepackage[latin9]{inputenc}
\usepackage{color}
\definecolor{note_fontcolor}{rgb}{0.80078125, 0.80078125, 0.80078125}
\usepackage[english]{babel}
\usepackage{amsmath}
\usepackage{amssymb}
\usepackage{graphicx}
\usepackage{esint}
\usepackage[unicode=true,
 bookmarks=true,bookmarksnumbered=true,bookmarksopen=true,bookmarksopenlevel=1,
 breaklinks=true,pdfborder={0 0 0},backref=false,colorlinks=true]
 {hyperref}
\hypersetup{
 pdfauthor={AINS}}

\makeatletter

\newenvironment{lyxgreyedout}
  {\textcolor{note_fontcolor}\bgroup\ignorespaces}
  {\ignorespacesafterend\egroup}

\@ifundefined{textcolor}{}
{%
 \definecolor{BLACK}{gray}{0}
 \definecolor{WHITE}{gray}{1}
 \definecolor{RED}{rgb}{1,0,0}
 \definecolor{GREEN}{rgb}{0,1,0}
 \definecolor{BLUE}{rgb}{0,0,1}
 \definecolor{CYAN}{cmyk}{1,0,0,0}
 \definecolor{MAGENTA}{cmyk}{0,1,0,0}
 \definecolor{YELLOW}{cmyk}{0,0,1,0}
 }
\numberwithin{equation}{section}
\numberwithin{figure}{section}
\numberwithin{table}{section}

\usepackage{graphicx}

\DeclareGraphicsExtensions{.png,.pdf,.eps,.jpg}
\graphicspath{{.}}

\@ifundefined{showcaptionsetup}{}{%
 \PassOptionsToPackage{caption=false}{subfig}}
\usepackage{subfig}
\makeatother

\begin{document}

\title{Relational causality and classical probability: Grounding quantum
phenomenology in a superclassical theory\vspace*{\bigskipamount}
}

\author{Gerhard \surname{Grössing}\textsuperscript{1}}

\email[E-mail: ]{ains@chello.at}

\homepage[Visit: ]{http://www.nonlinearstudies.at/}

\author{Siegfried \surname{Fussy}\textsuperscript{1}}

\email[E-mail: ]{ains@chello.at}

\homepage[Visit: ]{http://www.nonlinearstudies.at/}

\author{Johannes \surname{Mesa Pascasio}\textsuperscript{1,2}}

\email[E-mail: ]{ains@chello.at}

\homepage[Visit: ]{http://www.nonlinearstudies.at/}

\author{Herbert \surname{Schwabl}\textsuperscript{1}}

\email[E-mail: ]{ains@chello.at}

\homepage[Visit: ]{http://www.nonlinearstudies.at/}

\affiliation{\textsuperscript{1}Austrian Institute for Nonlinear Studies, Akademiehof\\
 Friedrichstr.~10, 1010 Vienna, Austria}

\affiliation{\textsuperscript{2}Institute for Atomic and Subatomic Physics, Vienna
University of Technology\\
Operng.~9, 1040 Vienna, Austria\vspace*{2cm}
}

\date{\today}
\begin{abstract}
By introducing the concepts of ``superclassicality'' and ``relational
causality'', it is shown here that the velocity field emerging from
an \emph{n}-slit system can be calculated as an average classical
velocity field with suitable weightings per channel. No deviation
from classical probability theory is necessary in order to arrive
at the resulting probability distributions. In addition, we can directly
show that when translating the thus obtained expression for said velocity
field into a more familiar quantum language, one immediately derives
the basic postulate of the de\,Broglie-Bohm theory, i.e.~the guidance
equation, and, as a corollary, the exact expression for the quantum
mechanical probability density current. Some other direct consequences
of this result will be discussed, such as an explanation of Born's
rule and Sorkin's first and higher order sum rules, respectively.
\begin{lyxgreyedout}
\noindent \global\long\def\VEC#1{\mathbf{#1}}
\global\long\def\d{\,\mathrm{d}}
\global\long\def\e{{\rm e}}
\global\long\def\meant#1{\left<#1\right>}
\global\long\def\meanx#1{\overline{#1}}
\global\long\def\mpbracket{\ensuremath{\genfrac{}{}{0pt}{1}{-}{\scriptstyle (\kern-1pt +\kern-1pt )}}}
\global\long\def\pmbracket{\ensuremath{\genfrac{}{}{0pt}{1}{+}{\scriptstyle (\kern-1pt -\kern-1pt )}}}
\global\long\def\p{\partial}
\end{lyxgreyedout}

\end{abstract}


\maketitle

\section{Introduction\label{sec:intro}}

In 1951, Richard Feynman published a paper \cite{Feynman.1951concept}
claiming that classical probability theory was not applicable for
the description of quantum phenomena, but that instead separate ``laws
of probabilities of quantum mechanics'' were required. In describing
the propagations of electrons from a source \emph{S} to a location
\emph{X} on the screen in a double-slit experiment, Feynman wrote:
\textquotedblleft{}We might at first suppose (since the electrons
behave as particles) that

I. Each electron which passes from \emph{S} to \emph{X} must go either
through hole 1 or hole 2. As a consequence of I we expect that:

II. The chance of arrival at X should be the sum of two parts, $P_{1}$,
the chance of arrival coming through hole 1, plus $P_{2}$, the chance
of arrival coming through hole 2.'' However, we apparently know from
experiment that this is not so (Fig.~1). Feynman concludes: \textquotedblleft{}Hence
experiment tells us definitely that $P\neq P_{1}+P_{2}$ or that II
is false. (\ldots{}) We must conclude that when both holes are open
it is not true that the particle goes through one hole or the other.''

\begin{figure}[h]
\centering{}\includegraphics[width=0.7\textwidth]{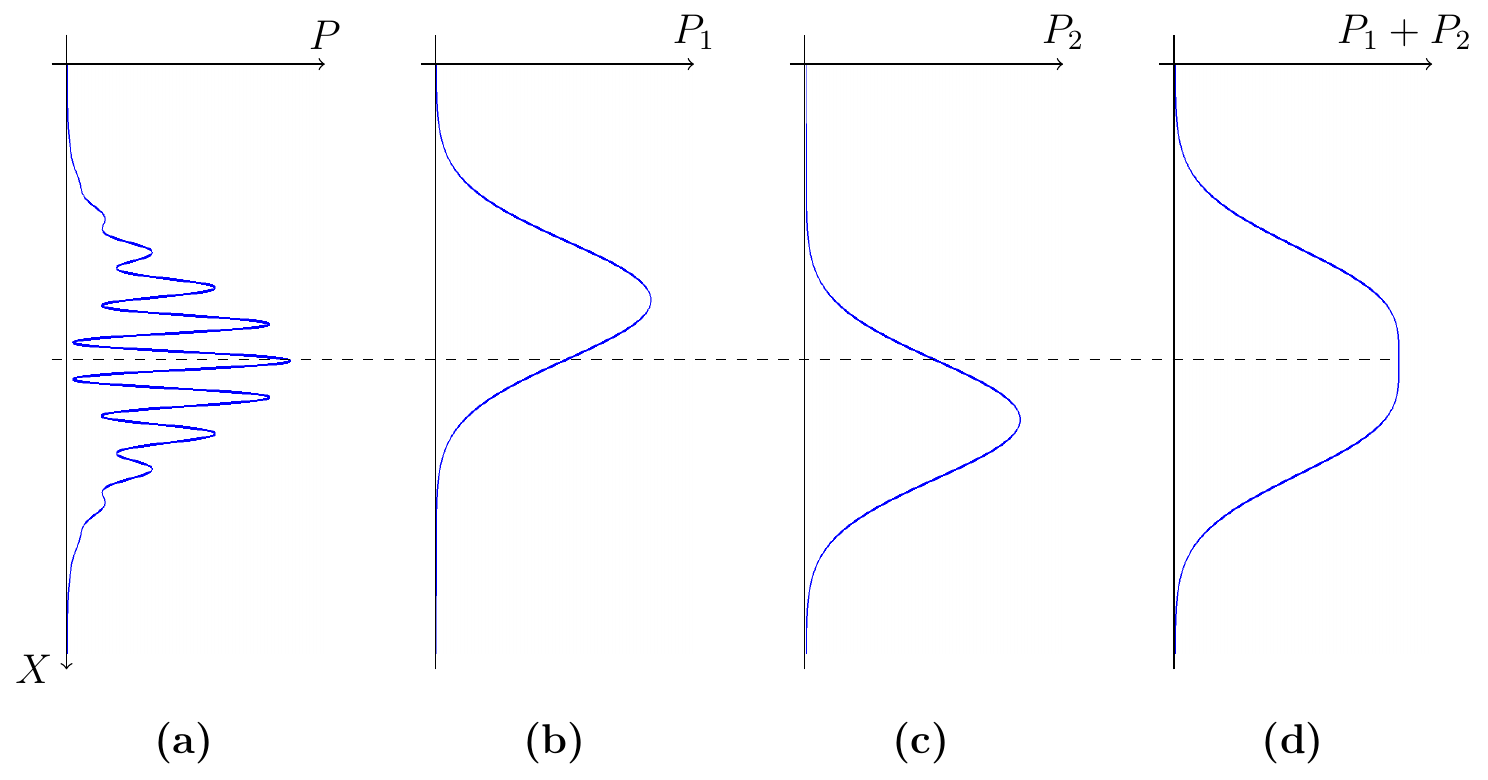}\caption{Scheme of interference at a double slit: probability distributions
for (a) both slits open: $P$, (b) slit 2 closed: $P_{1}$, (c) slit
1 closed: $P_{2}$, (d) the sum of (b) and (c): $P_{1}+P_{2}$.\label{fig:interf-1}}
\end{figure}

However, it was shown very clearly in several later papers, notably
by B.\,O.\ Koopman in 1955~\cite{Koopman.1955quantum} and by L.\,E.\ Ballentine
in 1986~\cite{Ballentine.1986probability}, that Feynman's reasoning
was not conclusive. In fact, Ballentine concluded that Feynman's ``argument
draws its radical conclusion from an incorrect application of probability
theory'', as one has to use conditional probabilities to account
for the different experimental contexts: $P_{1}(X|C_{1})$, $P_{2}(X|C_{2})$,
\emph{$P(X|C_{3})$}, where the contexts $C_{1}$, $C_{2}$ and $C_{3}$
refer to left, right and both slit(s) open, respectively. Then, it
is clear from the experimental data that $P(X|C_{3})\neq P_{1}(X|C_{1})+P_{2}(X|C_{2})$.
Accordingly, Ballentine also proved that ``the quantum mechanical
superposition principle for amplitudes is in no way incompatible with
the formalism of probability theory''~\cite{Ballentine.1986probability}.
Moreover, as one can easily show with the inclusion of the wave picture
for electrons (see Eq.~(\ref{eq:Sup1.3}) below), one correctly obtains
that $P(X|C_{3})=P_{1}(X|C_{3})+P_{2}(X|C_{3})$.

Note that the focus purely on context-free ``particles'' in quantum
mechanics has been the source of much confusion. We recall the clear
words of John Bell in this regard: ``While the founding fathers agonized
over the question `particle' or `wave', de\,Broglie in 1925 proposed
the obvious answer `particle' and `wave'. Is it not clear from the
smallness of the scintillation on the screen that we have to do with
a particle? And is it not clear, from the diffraction and interference
patterns, that the motion of the particle is directed by a wave? De\,Broglie
showed in detail how the motion of a particle, passing through just
one of two holes in a screen, could be influenced by waves propagating
through both holes. And so influenced that the particle does not go
where the waves cancel out, but is attracted to where they cooperate.
This idea seems to me so natural and simple, to resolve the wave-particle
dilemma in such a clear and ordinary way, that it is a great mystery
to me that it was so generally ignored.''~\cite{Bell.2004speakable}
Particularly since the impressive Paris experiments in the classical
domain performed by Yves Couder, Emmanuel Fort and co-workers~\cite{Couder.2006single-particle,Couder.2012probabilities,Fort.2010path-memory},
it is very suggestive that some sort of combination of ``particle''-
and wave-mechanics is at work also in the quantum domain. Essentially,
this amounts to obtaining particle distributions from calculating
classical-like wave intensity distributions. 

With wave-field intensities $P_{i}$ for each slit $i$ given by the
squared amplitude $R_{i},$ i.e.~with $P_{i}=R_{i}^{2}$, and with
the phase difference $\varphi,$ one obtains in the classical double
slit scenario for the intensity after slit 1
\begin{equation}
P\left(1\right)=P_{1}+R_{1}R_{2}\cos\varphi
\end{equation}
and for the intensity after slit 2
\begin{equation}
P\left(2\right)=P_{2}+R_{2}R_{1}\cos\varphi
\end{equation}
where the first expression on the r.h.s. refers to the intensity from
the slit \emph{per se} and the second expression refers to interference
with the other slit, respectively. This provides the total intensity
as 
\begin{equation}
P=P\left(1\right)+P\left(2\right)=P_{1}+P_{2}+2R_{1}R_{2}\cos\varphi.\label{eq:Sup1.3}
\end{equation}

In the following we try to shed more light on these issues by combining
results from the new field of \textquotedblleft{}Emergent Quantum
Mechanics\textquotedblright{}~\cite{Groessing.2012emerqum11-book}
with concepts of systems theory which we denote as ``relational causality''~\cite{Walleczek.2012mission}.
Since the physics of different scales is concerned, like, e.g., sub-quantum
and classical macro physics, we denote our sub-quantum theory as ``super-classical''.
(Both relational causality and superclassicality are to be defined
more specifically below in Section~\ref{sec:superclass}.) We consider
the quantum itself as an emergent system understood as off-equilibrium
steady state oscillation maintained by a constant throughput of energy
provided by the (\textquotedblleft{}classical\textquotedblleft{})
zero-point energy field. Starting with this concept, our group was
able to assess phenomena of standard quantum mechanics like Gaussian
dispersion of wave packets, superposition, double slit interference,
Planck's energy relation, or the Schrödinger equation, respectively,
as the emergent property of an underlying sub-structure of the vacuum
combined with diffusion processes reflecting the stochastic parts
of the zero-point field~\cite{Groessing.2008vacuum,Groessing.2009origin,Groessing.2012doubleslit,Groessing.2013dice}.

In Section~\ref{sec:guidance equation} we contrast the well-known
physics behind the double slit with an emergent vector field representation
of the observed interference field. The essential parts of our superclassical
approach are presented and the velocity field corresponding to the
guiding equation of the de\,Broglie-Bohm theory is derived. The explanation
and validity of Born's rule is analyzed in Section~\ref{sec:three-slit}
by means of a three slit configuration. In Section~\ref{sec:conclusion}
we summarize our results and give an outlook on possible limits of
the validity of present-day quantum mechanics from the perspective
of our sub-quantum approach.

\section{Introducing ``superclassicality'' and ``relational causality''\label{sec:superclass}}

In quantum mechanics, as well as in our quantum-like modeling via
an emergent quantum mechanics approach, one can write down a formula
for the total intensity distribution $P$ which is identical to (\ref{eq:Sup1.3}).
For the general case of $n$ slits, it holds with phase differences
$\varphi_{ij}=\varphi_{i}-\varphi_{j}$ that
\begin{equation}
P=\sum_{i=1}^{n}\left(P_{i}+\sum_{j=i+1}^{n}2R_{i}R_{j}\cos\varphi_{ij}\right),\label{eq:Sup2.1}
\end{equation}
where the phase differences are defined over the whole domain of the
experimental setup. Apart from the role of the relative phase with
important implications for the discussions on nonlocality~\cite{Groessing.2013dice},
there is one additional ingredient that distinguishes~(\ref{eq:Sup2.1})
from its classical counterpart~(\ref{eq:Sup1.3}), namely the ``dispersion
of the wavepacket''. As in our model the ``particle'' is actually
a ``bouncer'' in a fluctuating wave-like environment, i.e.~analogously
to the bouncers of Couder and Fort's group, one does have some (e.g.
Gaussian) distribution, with its center following the Ehrenfest trajectory
in the free case, but one also has a diffusion to the right and to
the left of the mean path which is just due to that stochastic bouncing.
Thus the total velocity field of our bouncer in its fluctuating environment
is given by the sum of the forward velocity $\VEC v$ and the respective
diffusive velocities $\VEC u_{L}$ and $\VEC u_{R}$ to the left and
the right. As for any direction $i$ the diffusion velocity $\VEC u_{i}=D\frac{\nabla_{i}P}{P}$
does not necessarily fall off with the distance, one has long effective
tails of the distributions which contribute to the nonlocal nature
of the interference phenomena~\cite{Groessing.2013dice}. In sum,
one has essentially three velocity (or current) channels per slit
in an $n-$slit system. 

Earlier we showed that this phenomenon can also be understood as a
variant of anomalous diffusion termed ``ballistic diffusion'': a
Brownian-type displacement with a time-dependent diffusivity (e.g.,
$D_{\mathrm{t}}=\frac{D^{2}}{\sigma_{0}^{2}}t$ in the case of a Gaussian
wave packet with standard deviation $\sigma$), leading to a ``classically''
obtained total velocity field
\begin{equation}
\mathbf{\mathrm{v}}_{\textrm{tot}}\left(t\right)=\mathbf{\mathrm{v}}\left(t\right)+\left[\mathbf{\mathrm{x}}_{\textrm{tot}}\left(t\right)-\mathrm{v}(t)\, t\right]\frac{D_{\mathrm{t}}}{\sigma^{2}},
\end{equation}
which is very practical to use in a computer simulation tool~\cite{Groessing.2011dice,Groessing.2010emergence}.
Moreover, ballistic diffusion is a signature of what has in recent
years become known as ``superstatistics''~\cite{Beck.2008}. Actually,
the prototype of a phenomenon amenable to superstatistics is just
a Brownian particle moving through a thermally changing environment:
Combining in one formalism a relatively fast dynamics (e.g. particle
velocity) on a small scale and slow changes (e.g.\ due to temperature
fluctuations) on a large scale leads to a superposition of two statistics,
i.e.~superstatistics. One of the main features of superstatistics
consists in \emph{emergent properties arising on intermediate scales},
which may be \emph{completely unexpected} if one looks only at either
the small or the large scale.

\begin{figure}
\centering{}\includegraphics[bb=0bp 10bp 705bp 420bp,clip,width=0.95\textwidth]{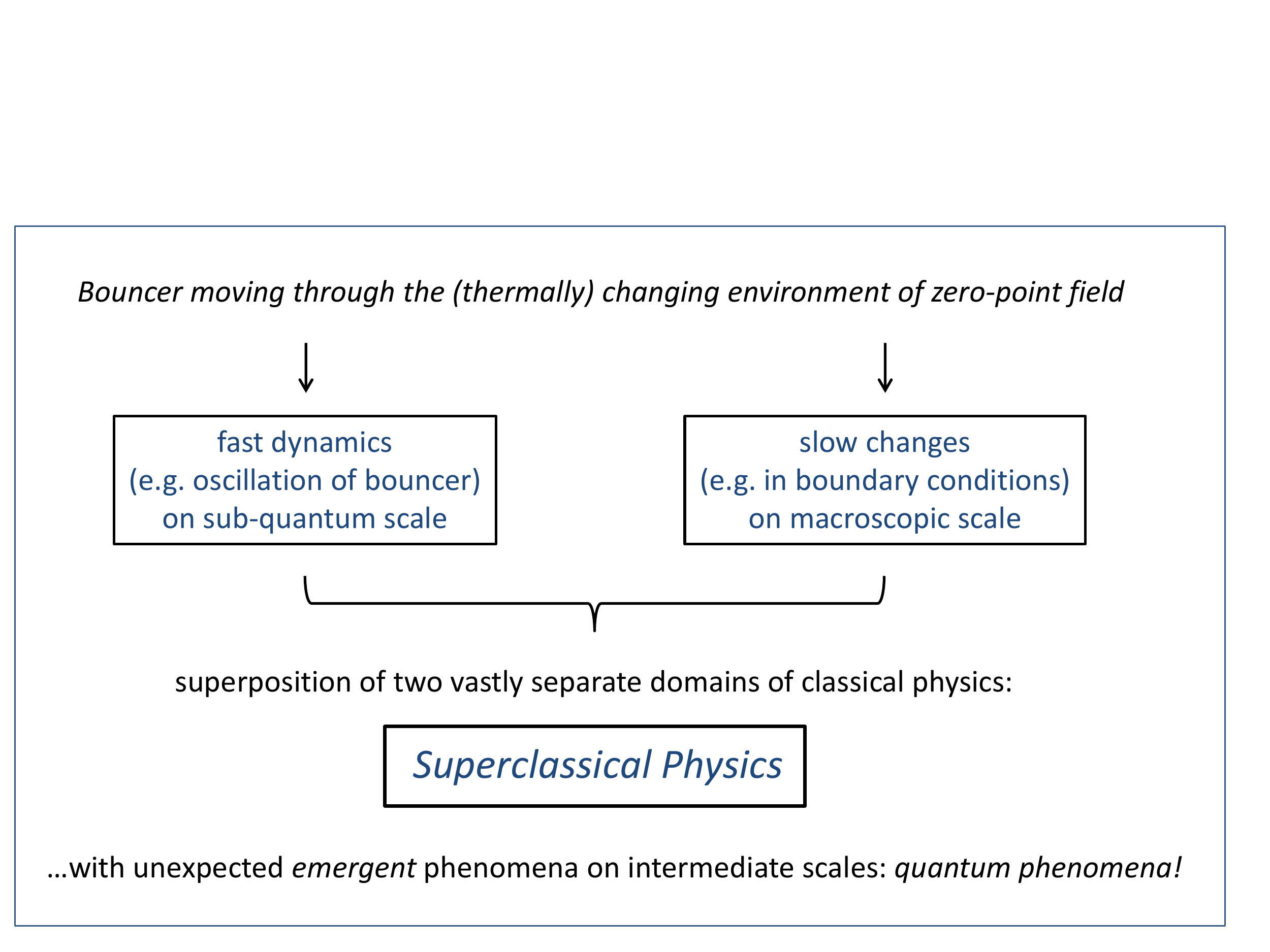}\caption{Scheme of \emph{Superclassical Physics} as applied to our bouncer
model\label{fig:superclass}}
\end{figure}

The phenomenon just described, however, also provides us with the
opportunity of solving a terminological problem with regard to sub-quantum
theories such as ours. For, although we consider our approach as based
on modern, ``21\textsuperscript{st} century classical physics'',
the latter must not be confused with the term ``classical physics''
as referring to the state-of-art of, say, the early 20\textsuperscript{th}
century, including e.g. the time of the first experimental evidence
of a zero-point energy (viz.~Mulliken's work from 1924~\cite{Mulliken.1924band},
which itself was before the advent of quantum theory proper). Acknowledging
both that a) a zero-point field can in principle considered a ``classical''
one and that b) quantum theory is characterized by some decisively
unexpected features when considered from a classical viewpoint, the
phenomenon of superstatistics thus suggests the following analogy
for the modeling of quantum systems: quantum behavior can be seen
to emerge from the interplay of classical processes at very small
(``sub-quantum'') and large (macroscopic) scales. A \emph{combination
of the physics on these vastly different scales} we call ``superclassical''
physics. The whole system under study then is characterized by \emph{processes
of emergence through the co-evolution of microscopic, local processes}
(like the oscillations of a bouncer) \emph{and of macroscopic processes}
(like the time evolution of the experimental boundary conditions).
In other words, starting from physical processes at such two vastly
different classical scales, their superposition (i.e.~within the
framework of superclassical physics) makes possible new types of phenomena
due to \emph{emergent features unexpected from either the very small
or the large scale physics} (Fig.~\ref{fig:superclass}; compare
also the related concept of ``emergent relativity'' as discussed,
e.g., by Jizba and Scardigli, this volume, \cite{Jizba.2012quantum},
and~\cite{Jizba.2012emergence}). In our development of an emergent
quantum mechanics, we shall thus further on describe it as a superclassical
approach, in order to avoid confusions when using ``classical''
explanations.

\begin{figure}
\centering{}\includegraphics[bb=0bp 80bp 720bp 540bp,clip,width=0.95\textwidth]{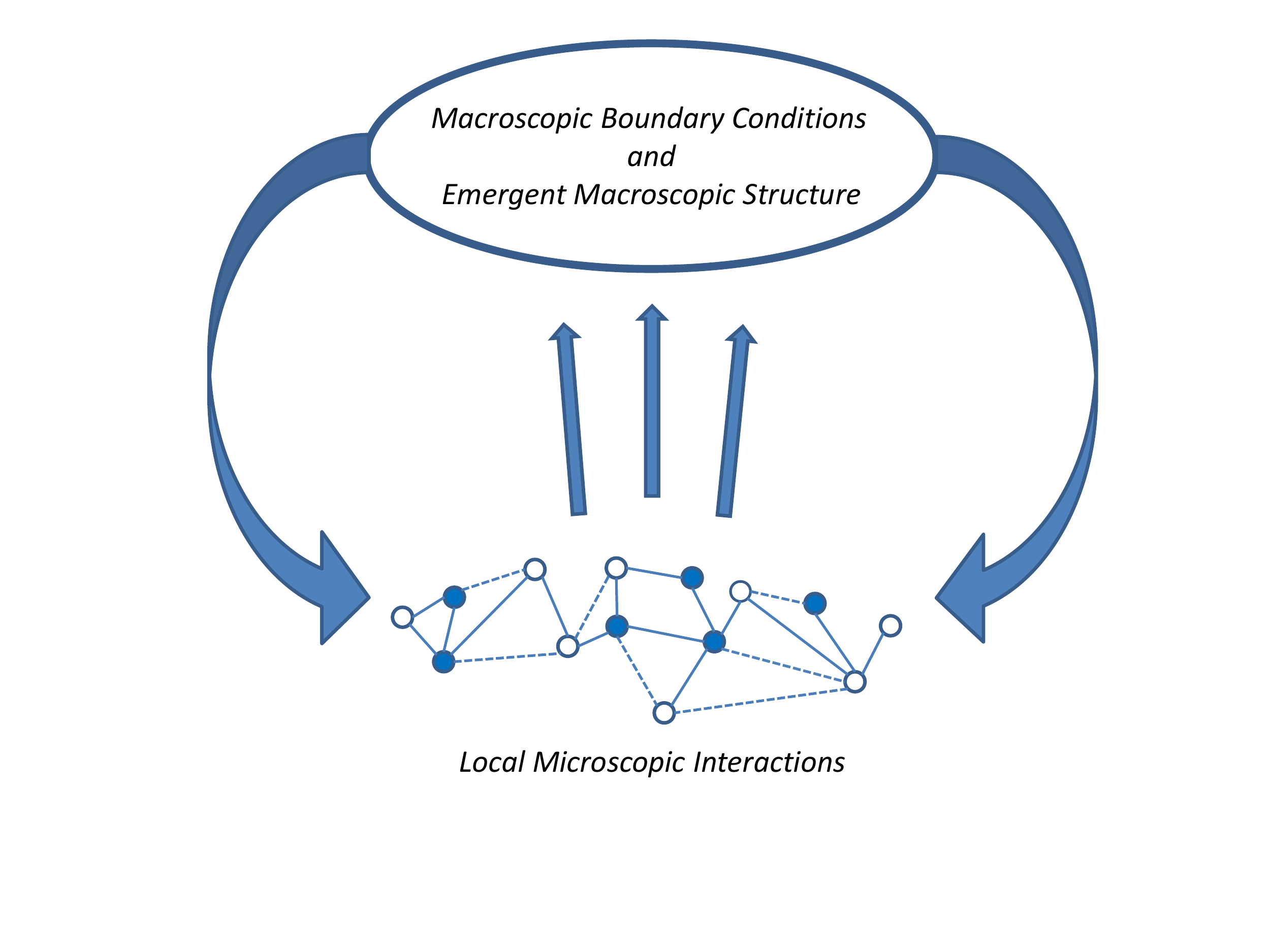}\caption{Scheme of \emph{Relational Causality}: Co-evolution of bottom-up and
top-down processes (after \cite{Walleczek.2000self-organized})\label{fig:relcaus}}
\end{figure}

As mentioned, our superclassical approach is particularly suited to
account for the emergent processes involved, i.e.~for processes of
a type that is well-known, for example, from the physics of Rayleigh-Bénard
cells. The latter appear in a fluid subjected to a temperature gradient
(like its container being heated from below) producing convection
rolls, with the emergent particle trajectories strongly depending
on the boundary conditions: The form of the convection rolls can be
radically changed by changing the boundaries of the container. Generally,
emergence of this type is characterized by \emph{relational causality},
i.e.~the co-evolution of bottom-up and top-down processes (Fig.~\ref{fig:relcaus}).

In our emergent quantum mechanics approach, we have given a superclassical
explanation of interference effects at a double slit~\cite{Groessing.2012doubleslit},
thereby arriving at expressions for the probability density current
and the corresponding velocity field completely equivalent to the
\emph{guidance equation}, which is the central postulate of the de\,Broglie-Bohm
interpretation~\cite{Bohm.1993undivided,Holland.1993}. Our claims,
to be substantiated in this paper, consist in the assertion that the
guidance equation is of some ``invisible hand'' type, i.e.\ somehow
mysteriously reaching out from configuration space in order to guide
particles in real three-space. Instead, we shall argue, the guidance
equation is completely understandable in real coordinate space, once
the concept of emergence is taken seriously and introduced within
a superclassical approach. 

\begin{figure}
\subfloat[Illustration of \emph{relational causality} under varying boundary
conditions as described in the text, with standing waves between plane
mirrors (black).\label{fig:relcaus-1a}]{\centering{}\includegraphics[width=0.95\textwidth]{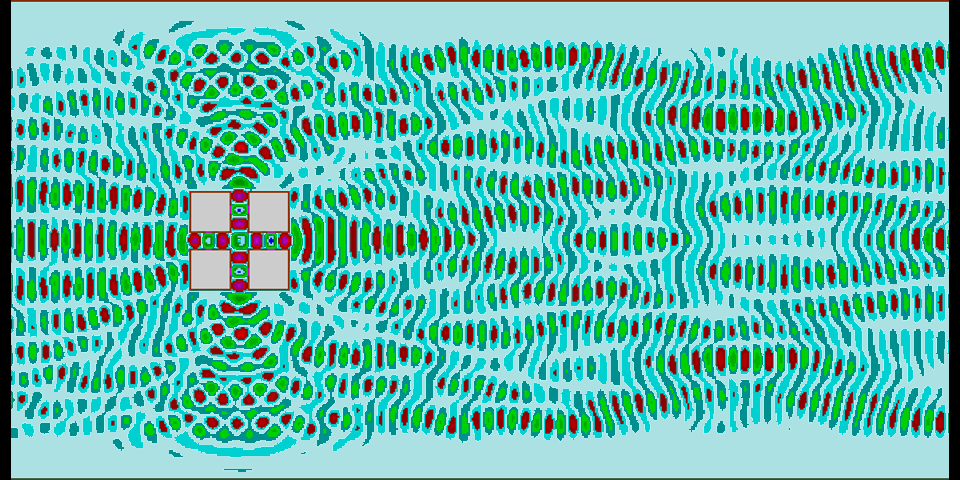}}

\subfloat[Same as (a), with modified standing waves due to modified geometry
of mirrors. Note the changes in the wave patterns around the squares
despite the fact that the local physics there is kept unchanged. These
changes are solely due to the altered boundary conditions.\label{fig:relcaus-1b}]{\centering{}\includegraphics[width=0.95\textwidth]{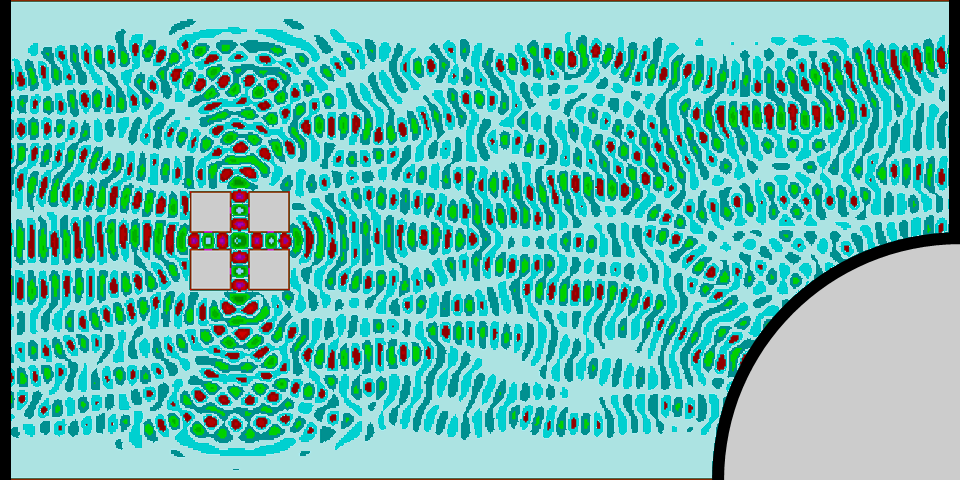}}

\caption{}
\end{figure}

Note that a basic characteristic of emergent systems can be described
via the above-mentioned relational causality (for a similar view,
see~\cite{Ellis.2012limits}). Consider for example the following
computer simulations (Figs.~\ref{fig:relcaus-1a} and \ref{fig:relcaus-1b}).
Similar to the experiments with droplets by Couder and Fort's group,
we insert onto a fluid surface a droplet in the center of the square
constituted by four smaller squares of solid material, such that due
to microscopic fluctuations the developing bouncer/walker will propagate
along one of the four narrow paths until it finally escapes from the
region of the square through one of the four slits into the open area.
With the bouncer creating waves that will be reflected from the walls
(drawn in black), after some time the whole system will develop standing
waves between said walls, which act as physically effective boundary
conditions. In Fig.~\ref{fig:relcaus-1a} the walls act as simple
plane mirrors, whereas in Fig.~\ref{fig:relcaus-1b} the reflecting
part on the right contains a circular structure, with the effect that
the pattern of standing waves is altered. The figures display intensity
distributions of the wave field which, accordingly, coincide with
probability density distributions of finding the droplet at a specific
location. Note that although the local physics in the vicinity of
the square would be identical for all four particle trajectories emerging
from one of the four slits \emph{if} one disregarded the context,
the pattern of the probability density field is different in both
cases. This is of course due to the fact that one cannot, in principle,
neglect the context, i.e.~the boundary conditions create different
standing wave patterns which in turn interfere with the otherwise
identical local processes. \emph{Relational causality} means that
one must consider the \emph{co-evolution of processes stemming from
the local slits and those from the global environment}, the latter
including the macroscopic boundary conditions. One can thus also speak
of a ``confluence'' of different currents, i.e.~those pertaining
to the four slits and those coming from the larger environment. Therefore,
if one records the bouncers' trajectories and collects them in a synoptic
manner, the whole velocity field will be a decisively \emph{emergent}
one. It is the totality of the whole experimental arrangement, and
not just the classically local influences, that results in the behavior
of the particle trajectories.

\begin{figure}
\centering{}\includegraphics{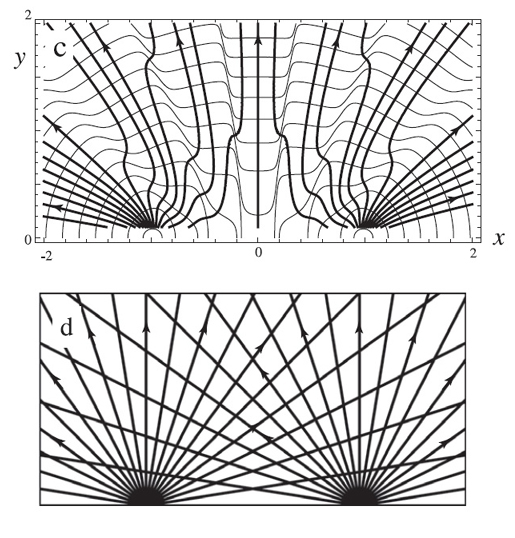}\caption{Quantum case streamlines (top) vs. geometrical rays (bottom). From
\cite{Berry.2009optical}. Here it is demonstrated that by bringing
together two sources, the temporal development of the velocity field
(and the corresponding ``trajectories'') in the quantum case can
be understood only if considered as the result of a complex interplay
of influences, updated at each instant of the time evolution, i.e.~as
the result of emergent processes.\label{fig:streamlines}}
\end{figure}

From this perspective, it is not very surprising that any relic of
context-free modeling should run into difficulties. Ironically, it
was an orthodox position that has argued against the consequences
of the more holistic approach acknowledging the context, i.e.~when
papers on apparently ``surreal trajectories'' were published against
the assertion of the de\,Broglie-Bohm model that particles would
not just follow trajectories as expected from context-free propagation
obeying simple classically-local momentum conservation (see, e.g.,
the assertions of Scully~\cite{Scully.1998bohm}). The latter would
apply only to geometrical rays, whereas the quantum case streamlines
are more complicated, a situation well-known from optics (see Fig.~\ref{fig:streamlines}
for illustration), and recently also confirmed via weak measurements
in a double-slit experiment involving photons~\cite{Kocsis.2011observing}.

Having thus introduced the notions of superclassical physics and relational
causality, we are now ready to reconsider interference at the double
slit. Recall that to arrive at the \emph{classical} double slit formula
for the intensities on the screen one had to just consider the mean
velocity $\mathrm{\mathbf{v}}$, i.e.~there was just one channel
(of the velocity, or the probability density current, respectively)
per slit. Because the currents represent wave propagations, this has
led to the total intensity of~(\ref{eq:Sup1.3}), $P=P\left(1\right)+P\left(2\right)=R_{1}^{2}+R_{2}^{2}+2R_{1}R_{2}\cos\varphi$.
Now, however, we are going to look for the \emph{superclassical} interference
formula. As mentioned, we now have to deal with three channels per
slit, due to the additional sub-quantum diffusion velocities $\VEC u_{L}$
and $\VEC u_{R}$, next to $\mathrm{\mathbf{v}}$, and \emph{all three
co-evolving}. Surprisingly, this is all that is needed to arrive at
the well-known quantum mechanical results which usually are obtained
only via complex-valued probability amplitudes.

\section{A superclassical derivation of the guidance equation\label{sec:guidance equation}}

Considering particles as oscillators (``bouncers'') coupling to
regular oscillations of the vacuum's zero-point field, which they
also generate, we have shown how a quantum can be understood as an
emergent system. In particular, with the dynamics between the oscillator
on the one hand, and the ``bath'' of its thermal environment as
constrained by the experiment's boundary conditions on the other hand,
one can explain not only Gaussian diffraction at a single slit~\cite{Groessing.2010emergence},
but also the well-known interference effects at a double slit~\cite{Groessing.2012doubleslit,Fussy.2014multislit}.
As already mentioned, we have also shown that the spreading of a wave
packet can be exactly described by combining the convective with the
orthogonal diffusive velocity fields. The latter fulfill the condition
of being unbiased w.r.t.\ the convective velocities, i.e.~the orthogonality
relation for the \textit{averaged} velocities derived in~\cite{Groessing.2010emergence}
is $\VEC{\overline{vu}}=0$, since any fluctuations $\VEC u=\delta\left(\nabla S/m\right)$
are shifts along the surfaces of action $\mathit{S=\mathrm{\mathrm{const}}.}$ 

To account for the different velocity channels $i=1,\ldots,3n$, $n$
being the number of slits, we now introduce for general cases the
generalized velocity vectors $\VEC w_{i}$, which in the case of $n=2$
are 
\begin{equation}
\VEC w_{1}:=\VEC v_{\mathrm{1}},\quad\VEC w_{2}:=\VEC u_{\mathrm{1R}},\quad\VEC w_{3}:=\VEC u_{\mathrm{1L}}\label{eq:nslit.2.4}
\end{equation}
for the first channel, and 
\begin{equation}
\VEC w_{4}:=\VEC v_{\mathrm{2}},\quad\VEC w_{5}:=\VEC u_{\mathrm{2R}},\quad\VEC w_{6}:=\VEC u_{\mathrm{2L}}\label{eq:nslit.2.5}
\end{equation}
for the second channel. The associated amplitudes $R(\VEC{w_{\mathrm{i}}})$
for each channel are taken to be the same, i.e.~$R(\VEC w_{1})=R(\VEC w_{2})=R(\VEC w_{3})=R_{1}$,
and $R(\VEC w_{4})=R(\VEC w_{5})=R(\VEC w_{6})=R_{2}$. 

Now, relational causality manifests itself in that the total wave
intensity field consists of the sum of all local intensities in each
channel, and the local intensity in each channel is the result of
the interference with the total intensity field. Thus, any change
in the local field affects the total field, and \emph{vice versa}:
Any change in the total field affects the local one. In order to completely
accommodate the totality of the system, we therefore need to define
a ``wholeness''-related local wave intensity $P(\VEC w_{i})$ in
one channel (i.e.~$\VEC w_{i}$) upon the condition that the totality
of the superposing waves is given by the ``rest'' of the $3n-1$
channels. Concretely, we account for the phase-dependent amplitude
contributions of the total system's wave field projected on one channel's
amplitude $R(\VEC w_{i})$ at the point $(\VEC x,t)$ with a \emph{conditional
probability density} $P(\VEC w_{i})$. The expression for $P(\VEC w_{i})$
thus constitutes the representation of relational causality within
our ansatz. Moreover, as usual one can define a local current $\mathbf{J}(\VEC w_{i})$
per channel as the corresponding ``local'' intensity-weighted velocity
$\VEC w_{i}$. Since the two-path set-up has $3n=6$ velocity vectors
at each point (cf.\ Eqs.~(\ref{eq:nslit.2.4}) and (\ref{eq:nslit.2.5})),
we thus obtain for the partial intensities and currents, respectively,
i.e.~for each channel component $\mathit{i}$,
\begin{align}
P(\VEC w_{i}) & =R(\VEC w_{i})\VEC{\hat{w}}_{i}\cdot{\displaystyle \sum_{j=1}^{6}}\VEC{\hat{w}}_{j}R(\VEC w_{j})\label{eq:Proj}\\
\VEC J\mathrm{(}\VEC w_{i}\mathrm{)} & =\VEC w_{i}P(\VEC w_{i}),\quad i=1,\ldots,6,
\end{align}
with
\begin{equation}
\cos\varphi_{i,j}:=\VEC{\hat{w}}_{i}\cdot\VEC{\hat{w}}_{j}.
\end{equation}
Consequently, the total intensity and current of our field read as
\begin{align}
P_{\mathrm{tot}}= & {\displaystyle \sum_{i=1}^{6}}P(\VEC w_{i})=\left({\displaystyle \sum_{i=1}^{6}}\VEC{\hat{w}}_{i}R(\VEC w_{i})\right)^{2}\label{eq:Ptot6}\\
\VEC J_{\mathrm{tot}}= & \sum_{i=1}^{6}\VEC J(\VEC w_{i})={\displaystyle \sum_{i=1}^{6}}\VEC w_{i}P(\VEC w_{i}),\label{eq:Jtot6}
\end{align}
 leading to the \textit{emergent total velocity}

\begin{equation}
\VEC v_{\mathrm{tot}}=\frac{\VEC J_{\mathrm{tot}}}{P_{\mathrm{tot}}}=\frac{{\displaystyle \sum_{i=1}^{6}}\VEC w_{i}P(\VEC w_{i})}{{\displaystyle \sum_{i=1}^{6}}P(\VEC w_{i})}\,.\label{eq:vtot_fin}
\end{equation}

Returning now to our previous notation for the six velocity components
$\VEC v_{i}$, $\VEC u_{i\mathrm{R}}$, $\VEC u_{i\mathrm{L}}$, $i=1,2$,
the partial current associated with $\VEC v_{\mathrm{1}}$ originates
from building the scalar product of $\VEC{\hat{v}}_{1}$ with all
other unit vector components and reads as
\begin{align}
\begin{array}{cc}
\VEC{J\mathrm{(}\VEC v_{\mathrm{1}})} & =\VEC v_{\mathrm{1}}P(\VEC v_{\mathrm{1}})=\VEC v_{\mathrm{1}}R_{1}\VEC{\hat{v}}_{1}\cdot(\VEC{\hat{v}}_{1}R_{1}+\VEC{\hat{u}}_{1\mathrm{R}}R_{1}+\VEC{\hat{u}}_{1\mathrm{L}}R_{1}+\VEC{\hat{v}}_{2}R_{2}+\VEC{\hat{u}}_{2\mathrm{R}}R_{2}+\VEC{\hat{u}}_{\mathrm{2L}}R_{2}).\end{array}\label{eq:Jv1}
\end{align}
Since trivially 
\begin{equation}
\VEC{\hat{u}}_{i\mathrm{R}}R_{i}+\VEC{\hat{u}}_{i\mathrm{L}}R_{i}=0,\quad i=1,2,\label{eq:triv}
\end{equation}
Eq.~(\ref{eq:Jv1}) leads to
\begin{align}
\VEC J\mathrm{(}\VEC v_{\mathrm{1}})=\VEC v_{\mathrm{1}}\left(R_{1}^{2}+R_{1}R_{2}\cos\varphi\right),
\end{align}
which results from the representation of the emerging velocity fields,
since we get the cosine of the phase difference $\varphi$ as a natural
result of the scalar product of the velocity vectors $\VEC v_{i}$.
The non-zero residua of the other vector fields yield 
\begin{equation}
\VEC J\mathrm{(}\VEC u_{\mathrm{1\mathrm{R}}}\mathrm{)}=u_{\mathrm{1R}}P\mathrm{(}\VEC u_{\mathrm{1\mathrm{R}}}\mathrm{)}=\VEC u_{\mathrm{1R}}\left(R_{1}\VEC{\hat{u}}_{\mathrm{1R}}\cdot\VEC{\hat{v}}_{2}R_{2}\right)=\VEC u_{\mathrm{1R}}R_{1}R_{2}\cos\left(\frac{\pi}{2}-\varphi\right)=\VEC u_{\mathrm{1R}}R_{1}R_{2}\sin\varphi
\end{equation}
and 
\begin{equation}
\VEC J\mathrm{(}\VEC{\VEC u_{\mathrm{1L}}})=\VEC u_{\mathrm{1L}}P\mathrm{(}\VEC u_{\mathrm{1L}}\mathrm{)}=\VEC u_{\mathrm{1L}}\left(R_{1}\VEC{\hat{u}}_{\mathrm{1L}}\cdot\VEC{\hat{v}}_{2}R_{2}\right)=\VEC u_{\mathrm{1L}}R_{1}R_{2}\cos\left(\frac{\pi}{2}+\varphi\right)=-\VEC u_{\mathrm{1L}}R_{1}R_{2}\sin\varphi.
\end{equation}
Analogously, we obtain for the convective velocity vector field of
the second channel
\begin{equation}
\VEC{J\mathrm{(}v_{\mathit{\mathrm{2}}}\mathrm{)}}=\VEC v_{\mathrm{2}}P(\VEC v_{\mathrm{2}})=\VEC v_{2}\left(R_{2}^{2}+R_{1}R_{2}\cos\varphi\right).
\end{equation}
The corresponding diffusive velocity vector fields read as
\begin{align}
\VEC J\mathrm{(}\VEC u_{\mathrm{2R}})= & \VEC u_{2R}P\mathrm{(}\VEC u_{2\mathrm{R}}\mathrm{)}=\VEC u_{2R}\left(R_{2}\VEC{\hat{u}}_{\mathrm{2R}}\cdot\VEC{\hat{v}}_{1}R_{1}\right)=\VEC u_{2R}R_{1}R_{2}\cos\left(\frac{\pi}{2}+\varphi\right)=-\VEC u_{2R}R_{1}R_{2}\sin\varphi,\\
\VEC J\mathrm{(}\VEC u_{\mathrm{2L}})= & \VEC u_{\mathrm{2L}}P\mathrm{(}\VEC u_{2\mathrm{L}}\mathrm{)}=\VEC u_{\mathrm{2L}}\left(R_{2}\VEC{\hat{u}}_{\mathrm{2L}}\cdot\VEC{\hat{v}}_{1}R_{1}\right)=\VEC u_{\mathrm{2L}}R_{1}R_{2}\cos\left(\frac{\pi}{2}-\varphi\right)=\VEC u_{\mathrm{2L}}R_{1}R_{2}\sin\varphi.
\end{align}
Note that the nontrivial sine contributions to the total current stem
from the projections between the diffusive velocities $\VEC u_{\mathrm{1R(L)}}$
of the first channel on the unit vector $\VEC{\hat{v}}_{2}$ of the
convective velocity of the second channel, and \textit{vice versa}.
Combining all terms, we obtain with Eq.~(\ref{eq:Jtot6}) the result
for the total current
\begin{align}
\VEC J_{\mathrm{tot}} & =\VEC v_{\mathrm{1}}P(\VEC v_{\mathrm{1}})+\VEC u_{1R}P\mathrm{(}\VEC u_{1\mathrm{R}}\mathrm{)}+\VEC u_{1\mathrm{L}}P\mathrm{(}\VEC u_{\mathrm{1\mathrm{L}}}\mathrm{)}+\VEC v_{\mathrm{2}}P(\VEC v_{\mathrm{2}})+\VEC u_{2R}P\mathrm{(}\VEC u_{2\mathrm{R}}\mathrm{)}+\VEC u_{\mathrm{2L}}P\mathrm{(}\VEC u_{2\mathrm{L}}\mathrm{)}\nonumber \\
 & =R_{1}^{2}\VEC v_{\mathrm{1}}+R_{2}^{2}\VEC v_{\mathrm{2}}+R_{1}R_{2}\left(\VEC v_{\mathrm{1}}+\VEC v_{\mathrm{2}}\right)\cos\varphi+R_{1}R_{2}\left([\VEC u_{1\mathrm{R}}-\VEC u_{1\mathrm{L}}]-[\VEC u_{2\mathrm{R}}-\VEC u_{2\mathrm{L}}]\right)\sin\varphi.\label{eq:Jtot}
\end{align}
The resulting diffusive velocities $\VEC u_{i\mathrm{\mathrm{R}}}-\VEC u_{i\mathrm{L}}$
are identified with the effective diffusive velocities $\VEC u_{i}$
for each channel. Note that \textit{one} of those velocities, $\VEC u_{i\mathrm{\mathrm{R}}}$
or $\VEC u_{i\mathrm{L}}$, respectively, is always zero, so that
the product of said difference with $\sin\varphi$ guarantees the
correct sign of the last term in Eq.~(\ref{eq:Jtot}). Thus we obtain
the final expression for the total density current built from the
remaining $2n=4$ velocity components
\begin{equation}
\VEC J_{\mathrm{tot}}=R_{1}^{2}\VEC v_{\mathrm{1}}+R_{2}^{2}\VEC v_{\mathrm{2}}+R_{1}R_{2}\left(\VEC v_{\mathrm{1}}+\VEC v_{2}\right)\cos\varphi+R_{1}R_{2}\left(\VEC u_{1}-\VEC u_{2}\right)\sin\varphi.\label{eq:Jfinal}
\end{equation}
The obtained total density current field $\VEC J_{\mathrm{tot}}(\VEC x,t)$
spanned by the various velocity components $\VEC v_{i}(\VEC x,t)$
and $\VEC u_{i\mathrm{R(L)}}(\VEC x,t)$ we have denoted as the ``path
excitation field'' \cite{Groessing.2012doubleslit}. It is built
by the sum of all its partial currents, which themselves are built
by an amplitude weighted projection of the total current.

Summing up the probabilities associated with each of the partial currents
we obtain according to the ansatz (\ref{eq:Proj}) and the relations
(\ref{eq:Ptot6}) and (\ref{eq:triv})
\begin{align}
P_{\mathrm{tot}} & =(R_{1}\VEC{\hat{v}}_{1}+R_{1}\VEC{\hat{u}}_{\mathrm{1R}}+R_{1}\VEC{\hat{u}}_{\mathrm{1L}}+R_{2}\VEC{\hat{v}}_{2}+R_{2}\VEC{\hat{u}}_{\mathrm{2R}}+R_{2}\VEC{\hat{u}}_{\mathrm{2L}})^{2}\nonumber \\
 & =(R_{1}\VEC{\hat{v}}_{1}+R_{2}\VEC{\hat{v}}_{2})^{2}=R_{1}^{2}+R_{2}^{2}+2R_{1}R_{2}\cos\varphi=P(\VEC v_{\mathrm{1}})+P(\VEC v_{\mathrm{2}}).\label{eq:Ptot2slit}
\end{align}
The total velocity $\VEC{v_{\mathrm{tot}}}$ according to Eq.~(\ref{eq:vtot_fin})
now reads as
\begin{equation}
\VEC v_{\mathrm{tot}}=\frac{R_{1}^{2}\VEC v_{\mathrm{1}}+R_{2}^{2}\VEC v_{\mathrm{2}}+R_{1}R_{2}\left(\VEC v_{\mathrm{1}}+\VEC v_{2}\right)\cos\varphi+R_{1}R_{2}\left(\VEC u_{1}-\VEC u_{2}\right)\sin\varphi}{R_{1}^{2}+R_{2}^{2}+2R_{1}R_{2}\cos\varphi}\,.\label{eq:vtot}
\end{equation}

The trajectories or streamlines, respectively, are obtained according
to $\VEC{\dot{x}}=\VEC v_{\mathrm{tot}}$ in the usual way by integration.
As first shown in \cite{Groessing.2012doubleslit}, by re-inserting
the expressions for convective and diffusive velocities, respectively,
i.e.~$\VEC v_{i,\mathrm{conv}}=\frac{\nabla S_{i}}{m}$, $\VEC u_{i}=-\frac{\hbar}{m}$$\frac{\nabla R_{i}}{R_{i}}$,
one immediately identifies Eq.~(\ref{eq:vtot}) with the Bohmian
guidance equation and Eq.~(\ref{eq:Jfinal}) with the quantum mechanical
pendant for the probability density current \cite{Sanz.2008trajectory}.
The latter can be seen as follows. Upon employment of the Madelung
transformation for each path $j$ ($j=1$ or $2$), 
\begin{equation}
\psi_{j}=R_{j}\e^{\mathrm{i}S_{j}/\hbar},\label{eq:3.14}
\end{equation}
and thus $P_{j}=R_{j}^{2}=|\psi_{j}|^{2}=\psi_{j}^{*}\psi_{j}$, with
$\varphi=(S_{1}-S_{2})/\hbar$, and recalling the usual trigonometric
identities such as $\cos\varphi=\frac{1}{2}\left(\e^{\mathrm{i}\varphi}+\e^{-\mathrm{i}\varphi}\right)$,
one can rewrite the total average current (\ref{eq:Jfinal}) immediately
as 
\begin{equation}
\begin{array}{rl}
{\displaystyle \mathbf{J}_{{\rm tot}}} & =P_{{\rm tot}}\mathbf{v}_{{\rm tot}}\\[3ex]
 & ={\displaystyle (\psi_{1}+\psi_{2})^{*}(\psi_{1}+\psi_{2})\frac{1}{2}\left[\frac{1}{m}\left(-\mathrm{i}\hbar\frac{\nabla(\psi_{1}+\psi_{2})}{(\psi_{1}+\psi_{2})}\right)+\frac{1}{m}\left(\mathrm{i}\hbar\frac{\nabla(\psi_{1}+\psi_{2})^{*}}{(\psi_{1}+\psi_{2})^{*}}\right)\right]}\\[3ex]
 & ={\displaystyle -\frac{\mathrm{i}\hbar}{2m}\left[\Psi^{*}\nabla\Psi-\Psi\nabla\Psi^{*}\right]={\displaystyle \frac{1}{m}{\rm Re}\left\{ \Psi^{*}(-\mathrm{i}\hbar\nabla)\Psi\right\} ,}}
\end{array}\label{eq:3.18}
\end{equation}
where $P_{{\rm tot}}=|\psi_{1}+\psi_{2}|^{2}=:|\Psi|^{2}$. The last
two expressions of (\ref{eq:3.18}) are the exact formulations of
the quantum mechanical probability current, here obtained just by
a re-formulation of (\ref{eq:Jfinal}). In fact, it is a simple exercise
to insert the wave functions (\ref{eq:3.14}) into (\ref{eq:3.18})
to re-obtain (\ref{eq:Jfinal}).

Note that it is straightforward to extend this derivation to the many-particle
case. As the individual terms in the expressions for the total current
and total probability density, respectively, are purely additive also
for \emph{N} particles, a fact that is well-known also from Bohmian
theory, the above-mentioned ``translation'' into orthodox quantum
language is straightforward, with the effect that the currents' nabla
operators just have to be applied at all of the locations $\mathbf{x}$
of the respective \emph{N} particles, thus providing the quantum mechanical
formula ${\displaystyle \mathbf{J}_{{\rm tot}}}\left(N\right)=\frac{1}{m}{\rm Re}\left\{ \Psi^{*}(-\mathrm{i}\hbar\nabla_{N})\Psi\right\} $,
where $\Psi$ now is the total \emph{N}-particle wave function. 

Again we emphasize that our result was obtained solely out of kinematic
relations by applying the superclassical rules introduced above on
the basis of a relational causality, i.e.~without invoking complex
$\psi$ functions or the like. Moreover, as opposed to the Bohmian
theory, we obtained our results not in configuration space but in
ordinary coordinate space. \textsl{What looks like the necessity to
superpose wave functions in configuration space, which then are imagined
to guide the particles by some invisible hand, can equally be obtained
by superpositions of all relational amplitude configurations of waves
in real space, i.e.\ by understanding the resulting system's evolutions
as processes of emergence.}

Thus, with $\VEC w_{i}=\frac{\VEC J(\VEC w_{i})}{P(\VEC w_{i})}$
and the classical composition principles of Eqs.~(\ref{eq:Ptot6})
and (\ref{eq:Jtot6}) we have shown that the total velocity field
is given in the simple form of a \emph{(super)classical average velocity
field}:

\begin{equation}
\VEC v_{\mathrm{tot}}=\frac{\VEC J_{\mathrm{tot}}}{P_{\mathrm{tot}}}=\frac{\sum_{i}\VEC J(\VEC w_{i})}{\sum_{i}P(\VEC w_{i})}=\frac{\sum_{i}\VEC w_{i}P(\VEC w_{i})}{\sum_{i}P(\VEC w_{i})}\,.
\end{equation}
In other words, the guidance equation postulated by de\,Broglie-Bohm
is here derived and explained via relational causality, with $\VEC v_{\mathrm{tot}}$
being an \emph{emergent velocity field}.

\section{Three-slit interference, Born's rule, and Sorkin's sum rules\label{sec:three-slit}}

The extension to three slits, beams, or probability current channels,
respectively, is straightforward. We just introduce a third emergent
propagation velocity $\VEC v_{3}$ and its corresponding diffusive
velocities $\VEC u_{3\mathrm{L(R)}}$. The phase shift of the third
beam is denoted as $\chi$ and represents the angle between the second
and the third beam in our geometric representation of the path excitation
field. According to Born's rule the probability of even a single particle
passing any of the three slits splits into a sum of probabilities
passing the slits pairwise, i.e.~going along both $\mathit{A}$ and
$\mathit{B}$, $\mathit{B}$ and $\mathit{C,}$ or $\mathit{A}$ and
$\mathit{C,}$ but never passing $\mathit{A}$, $\mathit{B}$ and
$\mathit{C}$ simultaneously. 

Interference phenomena have recently been analyzed thoroughly for
the cases of only one open slit up to $\mathit{n}$ open slits by
Sorkin~\cite{Sorkin.1994quantum}. For a double slit setup the interference
term is non-zero, i.e.~$I_{AB}:=P_{AB}-P_{A}-P_{B}\neq0$, with $P_{A(B)}$
being the detection probability with only one slit/path $\mathit{A}$
or $\mathit{B}$, respectively, of a total of $\mathit{n}$ slits/paths
open, and $P_{AB}$ for both slits $\mathit{A}$ and $\mathit{B}$
open. This ``first order sum rule'' is to be contrasted with Sorkin's
results for the following, so-called ``second order sum rule''~\cite{Sorkin.1994quantum}:
\begin{align}
I_{ABC}:= & P_{ABC}-P_{AB}-P_{AC}-P_{BC}+P_{A}+P_{B}+P_{C}\\
= & P_{ABC}-(P_{A}+P_{B}+P_{C}+I_{AB}+I_{AC}+I_{BC})=0.\nonumber 
\end{align}
This result is remarkable insofar as it can be inferred that interference
terms theoretically always originate from pairings of paths, but never
from triples etc. Any violation of this second order sum rule, i.e.~$I_{ABC}$$\neq0$,
and thus of Born's rule would have dramatic consequences for quantum
theory like a modification of the Schrödinger equation, for example.

Returning to our model, the total probability density current for
three paths is calculated according to the rules set up in Section~\ref{sec:guidance equation}.
We adopt the notations of the two slit system also for three slits,
i.e.~now employing nine velocity contributions: $\VEC v_{i}$, $\VEC u_{i\mathrm{R(L)}}$,
$i=1,2,3$. Analogously, the three generally different amplitudes
are denoted as $\mathrm{R}\mathrm{(}\VEC v_{i})=\mathrm{R}\mathrm{(}\VEC u_{i\mathrm{R}})=\mathrm{R}\mathrm{(}\VEC u_{i\mathrm{L}})=R_{i}$,
$i=1,2,3$. We keep the definition of $\varphi$ as $\varphi:=\arccos(\VEC{\hat{v}}_{1}$$\cdot$$\VEC{\hat{v}}_{2})$,
and we define the second angle as $\chi:=\arccos(\VEC{\hat{v}}_{2}\cdot\VEC{\hat{v}}_{3}).$
Similarly to Eq.~(\ref{eq:triv}), the diffusive velocities $\VEC u_{i\mathrm{\mathrm{R}}}-\VEC u_{i\mathrm{L}}$
combine to $\VEC u_{\mathrm{i}}$, $i=1,2,3$, thus ending up with
$2n=6$ effective velocities. Therefore we obtain, analogously to
the calculation in the previous section, 
\begin{align}
\VEC{J_{\mathrm{tot}}=} & R_{1}^{2}\VEC v_{\mathrm{1}}+R_{2}^{2}\VEC v_{\mathrm{2}}+R_{3}^{2}\VEC v_{\mathrm{3}}+R_{1}R_{2}\left(\VEC v_{\mathrm{1}}+\VEC v_{\mathrm{2}}\right)\cos\varphi+R_{1}R_{2}\left(\VEC u_{\mathrm{1}}-\VEC u_{\mathrm{2}}\right)\sin\varphi\nonumber \\
 & +R_{1}R_{3}\left(\VEC v_{\mathrm{1}}+\VEC v_{\mathrm{3}}\right)\cos\left(\varphi+\chi\right)+R_{1}R_{3}\left(\VEC u_{\mathrm{1}}-\VEC u_{\mathrm{3}}\right)\sin\left(\varphi+\chi\right)\nonumber \\
 & +R_{2}R_{3}\left(\VEC v_{\mathrm{2}}+\VEC v_{\mathrm{3}}\right)\cos\chi+R_{2}R_{3}\left(\VEC u_{\mathrm{2}}-\VEC u_{\mathrm{3}}\right)\sin\chi\label{eq:3.1}
\end{align}
and
\begin{align}
P_{\mathrm{tot}} & =R_{1}^{2}+R_{2}^{2}+R_{3}^{2}+2R_{1}R_{2}\cos\varphi+2R_{1}R_{3}\cos\left(\varphi+\chi\right)+2R_{2}R_{3}\cos\chi\label{eq:p3}\\
 & =P(\VEC v_{\mathrm{1}})+P(\VEC v_{\mathrm{2}})+P(\VEC v_{3}).\nonumber 
\end{align}

In analogy to the double slit case (cf.\ Eq.~(\ref{eq:Ptot2slit}))
we obtain a \textit{classical} Kolmogorov sum rule for the probabilities
on the one hand, but also the complete interference effects for the
double, three- and, as we have shown in~\cite{Fussy.2014multislit},
for the $n$-slit cases, on the other. However, the particular probabilities
$P(\VEC v_{i})$ in Eqs.~(\ref{eq:Ptot2slit}) and (\ref{eq:p3}),
do not correspond to the probabilities of the assigned slits if solely
opened, i.e.~$P_{AB}(\VEC v_{\mathrm{1}})=\left(R_{1}^{2}+R_{1}R_{2}\cos\varphi\right)\neq P_{A}(\VEC v_{\mathrm{1}})=R_{1}^{2}$.
Consequently, each of the probability summands in said equations does
\textit{not} correspond to an independent probability of the respective
slit if solely opened, a fact that was already clarified in our discussion
of the issue of contexts in Section \ref{sec:Introduction} .

Finally, we obtain for the cases of one (i.e.~$n=\mathit{A}$), two
and three open slits, respectively, 
\begin{equation}
I_{A}=P_{A}(\VEC v_{\mathrm{1}})=R_{1}^{2},
\end{equation}
\begin{equation}
I_{AB}=P_{AB}-P_{A}(\VEC v_{\mathrm{1}})-P_{B}(\VEC v_{\mathrm{2}})=2R_{1}R_{2}\cos\varphi,\label{eq:1storder}
\end{equation}
\begin{equation}
I_{ABC}=P_{ABC}-P_{AB}-P_{AC}-P_{BC}+P_{A}(\VEC v_{\mathrm{1}})+P_{B}(\VEC v_{\mathrm{2}})+P_{C}(\VEC v_{3})=0\,,\label{eq:2ndorder}
\end{equation}
where $P_{AB}$ is assigned to $P_{\mathrm{tot}}$ of Eq.~(\ref{eq:Ptot2slit})
and $P_{ABC}$ to $P_{\mathrm{tot}}$ of Eq.~(\ref{eq:p3}). In the
double slit case, e.g., with slits $\mathit{A}$ and $\mathit{B}$
open, we obtain the results of (\ref{eq:Ptot2slit}). If $\mathit{B}$
were closed and $\mathit{C}$ were open instead, we would get the
analogous result, i.e.~$\VEC v_{\mathrm{2}}$ and $\varphi$ replaced
by $\VEC v_{\mathrm{3}}$ and $\varphi_{1,3}$. If all three slits
$\mathit{A,B},C$ are open, we can use the pairwise permutations of
the double slit case, i.e.~$\mathit{A}\wedge B$, $A\wedge C$, or
$\mathit{B\wedge C}$, respectively, with $\varphi_{1,3}$ identified
with $\left(\varphi+\chi\right)$, etc. Thus we conclude that in our
model the addition of ``sub-probabilities'' indeed works and provides
the correct results.

Summarizing, with our superclassical model emerging out of a sub-quantum
scenario we arrive at the same results as standard quantum mechanics
fulfilling Sorkin's sum rules \cite{Sorkin.1994quantum}. However,
whereas in standard quantum mechanics Born's rule originates from
building the squared absolute values of additive $\psi$ functions
representing the probability amplitudes for different paths, in our
case we obtain the pairing of paths as a natural consequence of the
\textit{pairwise} selection of unit vectors of all existing velocity
components constituting the probability currents. Thus we obtain \textit{all}
possible pathways within an $\mathit{n}$-slit setup by our projection
method. The sum rules, Eqs.~(\ref{eq:Proj}) through (\ref{eq:vtot_fin}),
guarantee that each partial contribution, be it from the velocity
contributions within a particular channel or from different channels,
accounts for the final total current density for each point between
source and detector. Since for only one slit open the projection rule
(\ref{eq:Proj}) trivially leads to a linear relation between $\mathit{P}$
and $R^{2}$, the asymmetry between the latter quantities, due to
the nonlinear projection rule, becomes effective for $n\geq2$ slits
open. Consequently, the violation of the first order sum rule~(\ref{eq:1storder}),
i.e.~$I_{AB}\neq0$, represents a \textit{natural} result of our
principle of relational causality. Moreover, as we have argued above,
the opening of an additional slit solely adds pairwise path combinations.
As all higher interference terms have already incorporated said asymmetry,
the result can finally be reduced to the double slit case, thus yielding
a zero result as in Eq.~(\ref{eq:2ndorder}) according to Sorkin's
analysis.

This is a further hint that our model can reproduce all phenomena
of standard quantum theory with the option of giving a deeper reasoning
to principles like Born's rule or the hierarchical sum-rules, respectively.

\section{Conclusions and outlook\label{sec:conclusion}}

We have previously shown in a series of papers~\cite{Groessing.2008vacuum,Groessing.2009origin,Groessing.2012doubleslit,Groessing.2013dice,Groessing.2011dice,Groessing.2010emergence,Fussy.2014multislit,Groessing.2011explan}
that phenomena of standard quantum mechanics like Gaussian dispersion
of wave packets, superposition, double slit interference, Planck's
energy relation, or the Schrödinger equation can be assessed as the
emergent property of an underlying sub-structure of the vacuum combined
with diffusion processes reflecting also the stochastic parts of the
zero-point field, i.e.~the zero point fluctuations. (For similar
approaches see the works of Cetto and de la Pe\~na~\cite[and this volume]{Cetto.2012quantization},
Nieuwenhuizen~{[}this volume{]}, or Khrennikov\ \textsl{et\,al.}~\cite{Khrennikov.2012classical}.)
Thus we obtain the quantum mechanical dynamics as an averaged behavior
of sub-quantum processes. The inclusion of relativistic physics has
not been considered yet, but should be possible in principle. 

By introducing the concepts of superclassicality and relational causality,
we have in this paper shown that quantum phenomenology can be meaningfully
grounded in a superclassical approach relying solely on classical
probability theory. Apart from an application for a deeper understanding
of Born's rule, the central result of this work is a demonstration
that the guidance equation can be derived and explained within ordinary
coordinate space. We have proven the identity of our emergent velocity
field $\VEC v_{\mathrm{tot}}$ with the corresponding Bohmian one,
$\VEC v_{\mathrm{tot(Bohm)}}$, and the orthodox quantum mechanical
one, $\VEC v_{\mathrm{tot(QM)}}$, respectively: 

\begin{equation}
\begin{array}{rl}
{\displaystyle \mathbf{v}_{{\rm tot(emergent)}}} & =\frac{{\displaystyle \sum_{i}\VEC w_{i}P(\VEC w_{i})}}{{\displaystyle \sum_{i}P(\VEC w_{i})}}\,\\[3ex]
=\mathbf{v_{\mathsf{tot(Bohm)}}} & =\frac{\vphantom{{\displaystyle \oint}^{t}}{\displaystyle R_{1}^{2}\VEC v_{\mathrm{1}}+R_{2}^{2}\VEC v_{\mathrm{2}}+R_{1}R_{2}\left(\VEC v_{\mathrm{1}}+\VEC v_{2}\right)\cos\varphi+R_{1}R_{2}\left(\VEC u_{1}-\VEC u_{2}\right)\sin\varphi}}{{\displaystyle R_{1}^{2}+R_{2}^{2}+2R_{1}R_{2}\cos\varphi}}\\[3ex]
=\VEC v_{\mathrm{tot(QM)}} & ={\displaystyle {\displaystyle {\rm \frac{1}{m|\Psi|^{2}}Re}\left\{ \Psi^{*}(-\mathrm{i}\hbar\nabla)\Psi\right\} ,}}\quad\textrm{ with }\Psi=\sum_{j}\psi_{j}.
\end{array}
\end{equation}

\vspace{5mm}
Finally, with our superclassical theory we can also enquire into the
possible limits of present-day quantum theory. For example, the latter
is expected to break down at the time scales of our bouncer's oscillation
frequency, e.g., for the electron $\omega\approx\mathcal{O}\left(10^{21}\,\mathrm{Hz}\right)$.
As we have seen, at the emerging quantum level, i.e.~at times $t\ggg1/\omega$,
we obtain exact results strongly suggesting the validity of Born's
rule, for example. However, approaching said sub-quantum regions by
increasing the time resolution to the order of $t\approx1/\omega$
suggests a possibly gradual breakdown of said rule, since the averaging
of the diffusive and convective velocities and their mutual orthogonality
of the averaged velocities is not reliable any more. In principle,
this should eventually be testable in experiment. Moreover, upon the
velocities $\VEC v$ and $\VEC u_{\mathrm{L(R)}}$, introduction of
a new bias, either in the average orthogonality condition, or between
the different velocity channels, the question may be of relevance
whether these would lead to the collapse of the superposition principle,
as the assumed sub-quantum nonlinearities would then become manifest.
We have not discussed the important issue of nonlocality and possible
consequences with regard to the non-signaling principle here, and
refer the reader to the paper by Jan Walleczek (this volume) for consideration
of some of the topics in question.
\begin{acknowledgments}

\end{acknowledgments}
We thank Hans De\,Raedt for pointing out ref.~\cite{Ballentine.1986probability}
to us, Thomas Elze for the critical reading of an earlier version
of this paper, Jan Walleczek for many enlightening discussions, and
the Fetzer Franklin Fund for partial support of the current work.

\providecommand{\href}[2]{#2}\begingroup\raggedright
\endgroup


\begin{thebibliography}{10}

\bibitem{Feynman.1951concept}
R.~P. Feynman, ``The concept of probability in quantum mechanics,'' in {\em
  Proceedings of the Second Berkeley Symposium on Mathematical Statistics and
  Probability}, pp.~533--541.
\newblock University of California, Berkeley, {CA}, 1951.

\bibitem{Koopman.1955quantum}
B.~O. Koopman, ``Quantum theory and the foundations of probability,'' in {\em
  Applied Probability}, L.~A. {MacColl}, ed., pp.~97--102.
\newblock {McGraw-Hill}, New York, 1955.

\bibitem{Ballentine.1986probability}
L.~E. Ballentine, ``Probability theory in quantum mechanics,''
  \href{http://dx.doi.org/10.1119/1.14783}{{\em Am. J. Phys.} {\bfseries 54}
  (1986) 883}.

\bibitem{Bell.2004speakable}
J.~S. Bell, {\em Speakable and Unspeakable in Quantum Mechanics}.
\newblock Cambridge University Press, Cambridge, 2~ed., 2004.

\bibitem{Couder.2006single-particle}
Y.~Couder and E.~Fort, ``Single-particle diffraction and interference at a
  macroscopic scale,''
  \href{http://dx.doi.org/10.1103/PhysRevLett.97.154101}{{\em Phys. Rev. Lett.}
  {\bfseries 97} (2006) 154101}.

\bibitem{Couder.2012probabilities}
Y.~Couder and E.~Fort, ``Probabilities and trajectories in a classical
  wave-particle duality,''
  \href{http://dx.doi.org/10.1088/1742-6596/361/1/012001}{{\em J. Phys.: Conf.
  Ser.} {\bfseries 361} (2012) 012001}.

\bibitem{Fort.2010path-memory}
E.~Fort, A.~Eddi, A.~Boudaoud, J.~Moukhtar, and Y.~Couder, ``Path-memory
  induced quantization of classical orbits,''
  \href{http://dx.doi.org/10.1073/pnas.1007386107}{{\em {PNAS}} {\bfseries 107}
  (2010) 17515--17520}.

\bibitem{Groessing.2012emerqum11-book}
G.~Gr{\"o}ssing, ed., {\em Emergent Quantum Mechanics 2011}.
\newblock No.~361/1. {IOP} Publishing, Bristol, 2012.
\newblock http://iopscience.iop.org/1742-6596/361/1.

\bibitem{Walleczek.2012mission}
J.~Walleczek, ``Mission statement,'' 2012.
\newblock \url{http://www.fetzer-franklin-fund.org/mission/}.

\bibitem{Groessing.2008vacuum}
G.~Gr{\"o}ssing, ``The vacuum fluctuation theorem: Exact {S}chr{\"o}dinger
  equation via nonequilibrium thermodynamics,''
  \href{http://dx.doi.org/10.1016/j.physleta.2008.05.007}{{\em Phys. Lett. A}
  {\bfseries 372} (2008) 4556--4563},
  \href{http://arxiv.org/abs/0711.4954}{{\ttfamily {arXiv}:0711.4954
  [quant-ph]}}.

\bibitem{Groessing.2009origin}
G.~Gr{\"o}ssing, ``On the thermodynamic origin of the quantum potential,''
  \href{http://dx.doi.org/10.1016/j.physa.2008.11.033}{{\em Physica A}
  {\bfseries 388} (2009) 811--823},
  \href{http://arxiv.org/abs/0808.3539}{{\ttfamily {arXiv}:0808.3539
  [quant-ph]}}.

\bibitem{Groessing.2012doubleslit}
G.~Gr{\"o}ssing, S.~Fussy, J.~Mesa~Pascasio, and H.~Schwabl, ``An explanation
  of interference effects in the double slit experiment: Classical trajectories
  plus ballistic diffusion caused by zero-point fluctuations,''
  \href{http://dx.doi.org/10.1016/j.aop.2011.11.010}{{\em Ann. Phys.}
  {\bfseries 327} (2012) 421--437},
  \href{http://arxiv.org/abs/1106.5994}{{\ttfamily {arXiv}:1106.5994
  [quant-ph]}}.

\bibitem{Groessing.2013dice}
G.~Gr{\"o}ssing, S.~Fussy, J.~Mesa~Pascasio, and H.~Schwabl, ``{'Systemic}
  nonlocality' from changing constraints on sub-quantum kinematics,''
  \href{http://dx.doi.org/10.1088/1742-6596/442/1/012012}{{\em J. Phys.: Conf.
  Ser.} {\bfseries 442} (2013) 012012},
  \href{http://arxiv.org/abs/1303.2867}{{\ttfamily {arXiv}:1303.2867
  [quant-ph]}}.

\bibitem{Groessing.2011dice}
G.~Gr{\"o}ssing, S.~Fussy, J.~Mesa~Pascasio, and H.~Schwabl, ``Elements of
  sub-quantum thermodynamics: Quantum motion as ballistic diffusion,''
  \href{http://dx.doi.org/10.1088/1742-6596/306/1/012046}{{\em J. Phys.: Conf.
  Ser.} {\bfseries 306} (2011) 012046},
  \href{http://arxiv.org/abs/1005.1058}{{\ttfamily {arXiv}:1005.1058
  [physics.gen-ph]}}.

\bibitem{Groessing.2010emergence}
G.~Gr{\"o}ssing, S.~Fussy, J.~Mesa~Pascasio, and H.~Schwabl, ``Emergence and
  collapse of quantum mechanical superposition: Orthogonality of reversible
  dynamics and irreversible diffusion,''
  \href{http://dx.doi.org/10.1016/j.physa.2010.07.017}{{\em Physica A}
  {\bfseries 389} (2010) 4473--4484},
  \href{http://arxiv.org/abs/1004.4596}{{\ttfamily {arXiv}:1004.4596
  [quant-ph]}}.

\bibitem{Beck.2008}
C.~Beck, ``Superstatistics: Theoretical concepts and physical applications,''
  in {\em Anomalous Transport: Foundations and Applications}, R.~Klages,
  G.~Radons, and I.~M. Sokolov, eds., pp.~433--457.
\newblock Wiley-{VCH}, 2008.

\bibitem{Mulliken.1924band}
R.~S. Mulliken, ``The band spectrum of boron monoxide,''
  \href{http://dx.doi.org/10.1038/114349a0}{{\em Nature} {\bfseries 114} (1924)
  349--350}.

\bibitem{Jizba.2012quantum}
P.~Jizba and F.~Scardigli, ``Quantum mechanics and local {L}orentz symmetry
  violation,'' \href{http://dx.doi.org/10.1088/1742-6596/361/1/012026}{{\em J.
  Phys.: Conf. Ser.} {\bfseries 361} (2012) 012026}.

\bibitem{Jizba.2012emergence}
P.~Jizba and F.~Scardigli, ``Emergence of special and doubly special
  relativity,'' \href{http://dx.doi.org/10.1103/PhysRevD.86.025029}{{\em Phys.
  Rev. D} {\bfseries 86} (2012) 025029},
  \href{http://arxiv.org/abs/1105.3930}{{\ttfamily {arXiv}:1105.3930
  [hep-th]}}.

\bibitem{Walleczek.2000self-organized}
J.~Walleczek, ed., {\em Self-Organized Biological Dynamics and Nonlinear
  Control}.
\newblock Cambridge University Press, New York, 2000.

\bibitem{Bohm.1993undivided}
D.~Bohm and B.~J. Hiley, {\em The Undivided Universe: An Ontological
  Interpretation of Quantum Theory}.
\newblock Routledge, London, 1993.

\bibitem{Holland.1993}
P.~R. Holland, {\em The Quantum Theory of Motion}.
\newblock Cambridge University Press, Cambridge, 1993.

\bibitem{Ellis.2012limits}
G.~F. Ellis, ``On the limits of quantum theory: Contextuality and the
  quantum-classical cut,''
  \href{http://dx.doi.org/10.1016/j.aop.2012.05.002}{{\em Ann. Phys.}
  {\bfseries 327} (2012) 1890--1932},
  \href{http://arxiv.org/abs/1108.5261}{{\ttfamily {arXiv}:1108.5261
  [quant-ph]}}.

\bibitem{Berry.2009optical}
M.~V. Berry, ``Optical currents,''
  \href{http://dx.doi.org/10.1088/1464-4258/11/9/094001}{{\em J. Opt. A: Pure
  Appl. Opt.} {\bfseries 11} (2009) 094001}.

\bibitem{Scully.1998bohm}
M.~O. Scully, ``Do {B}ohm trajectories always provide a trustworthy physical
  picture of particle motion?,''
  \href{http://dx.doi.org/10.1238/Physica.Topical.076a00041}{{\em Phys. Scr.}
  {\bfseries T76} (1998) 41--46}.

\bibitem{Kocsis.2011observing}
S.~Kocsis, B.~Braverman, S.~Ravets, M.~J. Stevens, R.~P. Mirin, L.~K. Shalm,
  and A.~M. Steinberg, ``Observing the average trajectories of single photons
  in a two-slit interferometer,''
  \href{http://dx.doi.org/10.1126/science.1202218}{{\em Science} {\bfseries
  332} (2011) 1170--1173}.

\bibitem{Fussy.2014multislit}
S.~Fussy, J.~Mesa~Pascasio, H.~Schwabl, and G.~Gr{\"o}ssing, ``Born's rule as
  signature of a superclassical current algebra,''
  \href{http://dx.doi.org/10.1016/j.aop.2014.02.002}{{\em Ann. Phys.}
  {\bfseries 343} (2014) 200--214},
  \href{http://arxiv.org/abs/1308.5924}{{\ttfamily {arXiv}:1308.5924
  [quant-ph]}}.

\bibitem{Sanz.2008trajectory}
{\'A}.~S. Sanz and S.~Miret-Art{\'e}s, ``A trajectory-based understanding of
  quantum interference,''
  \href{http://dx.doi.org/10.1088/1751-8113/41/43/435303}{{\em J. Phys. A:
  Math. Gen.} {\bfseries 41} (2008) 435303},
  \href{http://arxiv.org/abs/0806.2105}{{\ttfamily {arXiv}:0806.2105
  [quant-ph]}}.

\bibitem{Sorkin.1994quantum}
R.~D. Sorkin, ``Quantum mechanics as quantum measure theory,''
  \href{http://dx.doi.org/10.1142/S021773239400294X}{{\em Mod. Phys. Lett. A}
  {\bfseries 09} (1994) 3119--3127},
  \href{http://arxiv.org/abs/gr-qc/9401003}{{\ttfamily {arXiv}:gr-qc/9401003}}.

\bibitem{Groessing.2011explan}
G.~Gr{\"o}ssing, J.~Mesa~Pascasio, and H.~Schwabl, ``A classical explanation of
  quantization,'' \href{http://dx.doi.org/10.1007/s10701-011-9556-1}{{\em
  Found. Phys.} {\bfseries 41} (2011) 1437--1453},
  \href{http://arxiv.org/abs/0812.3561}{{\ttfamily {arXiv}:0812.3561
  [quant-ph]}}.

\bibitem{Cetto.2012quantization}
A.~M. Cetto and L.~de~la Pe{\~n}a, ``Quantization as an emergent phenomenon due
  to matter-zeropoint field interaction,''
  \href{http://dx.doi.org/10.1088/1742-6596/361/1/012013}{{\em J. Phys.: Conf.
  Ser.} {\bfseries 361} (2012) 012013}.

\bibitem{Khrennikov.2012classical}
A.~Khrennikov, B.~Nilsson, and S.~Nordebo, ``Classical signal model reproducing
  quantum probabilities for single and coincidence detections,''
  \href{http://dx.doi.org/10.1088/1742-6596/361/1/012030}{{\em J. Phys.: Conf.
  Ser.} {\bfseries 361} (2012) 012030}.

\end{thebibliography}
\end{document}